\def\etc{{\it etc.}}

\def\~{{$\tilde{\phantom{a}}$}}

\documentclass [12pt] {article}
\usepackage{epsfig}
\usepackage{color}

\def\thebibliography#1{\section{References}\markboth
 {REFERENCES}{REFERENCES}\list
 {[\arabic{enumi}]}{\settowidth\labelwidth{[#1]}\leftmargin\labelwidth
 \advance\leftmargin\labelsep
 \usecounter{enumi}}
 \def\newblock{\hskip .11em plus .33em minus -.07em}
 \sloppy
 \sfcode`\.=1000\relax}
\def\upcite#1{\raise6pt\hbox{\scriptsize
\cite{#1}}}
\def\lsim{\mathrel {\vcenter {\baselineskip 0pt \kern 0pt
    \hbox{$<$} \kern 0pt \hbox{$\sim$} }}}
\def\gsim{\mathrel {\vcenter {\baselineskip 0pt \kern 0pt
    \hbox{$>$} \kern 0pt \hbox{$\sim$} }}}
\def\gtlt{\mathrel {\vcenter {\baselineskip 0pt \kern 0pt
    \hbox{$>$} \kern 0pt \hbox{$<$} }}}

\def\hline{\noalign{\hrule \vskip2pt}}

\def\|{\ifmmode\Vert\else \char`\|\fi}
\ifx\oldzeta\undefined                          
  \let\oldzeta=\zeta                            
  \def\zzeta{{\raise 2pt\hbox{$\oldzeta$}}}     
  \let\zeta=\zzeta                              
\fi

\ifx\oldchi\undefined                           
  \let\oldchi=\chi                              
  \def\cchi{{\raise 2pt\hbox{$\oldchi$}}}       
  \let\chi=\cchi                                
\fi

\def\frac#1#2{{#1 \over #2}}

\def\half{\ifinner {\scriptstyle {1 \over 2}}
   \else {1 \over 2} \fi}

\def\abs#1{\left\vert#1\right\vert}	

\def\simge{\mathrel{%
   \rlap{\raise 0.511ex \hbox{$>$}}{\lower 0.511ex \hbox{$\sim$}}}}
\def\simle{\mathrel{
   \rlap{\raise 0.511ex \hbox{$<$}}{\lower 0.511ex \hbox{$\sim$}}}}

\def\buildchar#1#2#3{{\null\!                   
   \mathop#1\limits^{#2}_{#3}                   
   \!\null}}                                    
\def\overcirc#1{\buildchar{#1}{\circ}{}}

\def\slashchar#1{\setbox0=\hbox{$#1$}           
   \dimen0=\wd0                                 
   \setbox1=\hbox{/} \dimen1=\wd1               
   \ifdim\dimen0>\dimen1                        
      \rlap{\hbox to \dimen0{\hfil/\hfil}}      
      #1                                        
   \else                                        
      \rlap{\hbox to \dimen1{\hfil$#1$\hfil}}   
      /                                         
   \fi}                                         %

\def\subrightarrow#1{
  \setbox0=\hbox{
    $\displaystyle\mathop{}
    \limits_{#1}$}
  \dimen0=\wd0
  \advance \dimen0 by .5em
  \mathrel{
    \mathop{\hbox to \dimen0{\rightarrowfill}}
       \limits_{#1}}}                           

\def\overlay#1#2{\ifmmode%
\setbox0=\hbox{$#1$}%
\setbox1=\hbox to\wd0{\hss$#2$\hss}\else%
\setbox0=\hbox{#1}%
\setbox1=\hbox to\wd0{\hss#2\hss}\fi%
#1\hskip-\wd0\box1 }

\def\pmb#1{\leavevmode\setbox0=\hbox{#1}%
\kern-.02em\copy0\kern-\wd0
\kern.04em\copy0\kern-\wd0
\kern-.02em\raise.04em\box0 }

\def\vereq#1#2{\lower3pt\vbox{\baselineskip1.5pt \lineskip1.5pt
\ialign{$\m@th#1\hfill##\hfil$\crcr#2\crcr\sim\crcr}}}

\def\tensor#1{\protect\@ontopof{#1}{\leftrightarrow}{1.15}\mathord{\box2}}
\def\overstar#1{\protect\@ontopof{#1}{\ast}{1.15}\mathord{\box2}}
\def\overdots#1{\protect\@ontopof{#1}{\cdots}{1.0}\mathord{\box2}}
\def\overcirc#1{\protect\@ontopof{#1}{\circ}{1.2}\mathord{\box2}}
\def\loarrow#1{\protect\@ontopof{#1}{\leftarrow}{1.15}\mathord{\box2}}
\def\roarrow#1{\protect\@ontopof{#1}{\rightarrow}{1.15}\mathord{\box2}}

\def\@ontopof#1#2#3{%
{\mathchoice
{\@@ontopof{#1}{#2}{#3}\displaystyle\scriptstyle}%
{\@@ontopof{#1}{#2}{#3}\textstyle\scriptstyle}%
{\@@ontopof{#1}{#2}{#3}\scriptstyle\scriptscriptstyle}%
{\@@ontopof{#1}{#2}{#3}\scriptscriptstyle\scriptscriptstyle}%
}%
}

\def\@@ontopof#1#2#3#4#5{%
\setbox0=\hbox{$#4#1$}%
\setbox1=\hbox{$#5#2$}%
\setbox2=\hbox{}\ht2=\ht0 \dp2=\dp0 %
\ifdim\wd0>\wd1 %
\setbox1=\hbox to\wd0{\hss\box1\hss}%
\mathord{\rlap{\raise#3\ht0\box1}\box0}%
\else   %
\setbox1=\hbox to.9\wd1{\hss\box1\hss}%
\setbox0=\hbox to\wd1{\hss$#4\relax#1$\hss}%
\mathord{\rlap{\copy0}\raise#3\ht0\box1}%
\fi
}%

\def\lambdabar{\protect\@lambdabar}
\def\@lambdabar{%
\relax
\bgroup
\def\@tempa{\hbox{\raise.73\ht0
\hbox to0pt{\kern.25\wd0\vrule width.5\wd0
height.1pt depth.1pt\hss}\box0}}%
\mathchoice{\setbox0\hbox{$\displaystyle\lambda$}\@tempa}%
{\setbox0\hbox{$\textstyle\lambda$}\@tempa}%
{\setbox0\hbox{$\scriptstyle\lambda$}\@tempa}%
{\setbox0\hbox{$\scriptscriptstyle\lambda$}\@tempa}%
\egroup
}

\def\corresponds{{\lower.2ex\hbox{=}}{\rm\kern-.75em^\triangle}}
\def\succsim{\succ\kern-.9em_\sim\kern.3em}
\def\precsim{\prec\kern-1em_\sim\kern.3em}
\def\slantfrac#1#2{\kern1em^{#1}\kern-.3em/\kern-.1em_{#2}}

\begin{document}
                                                                
\begin{center}
{\Large\bf A Conducting Checkerboard}
\\

\medskip

Kirk T.~McDonald
\\
{\sl Joseph Henry Laboratories, Princeton University, Princeton, NJ 08544}
\\
(October 4, 2001)
\end{center}

\section{Problem}

Some biological systems  consist of two ``phases" 
of nearly square fiber bundles of differing thermal and electrical conductivities.  Consider a circular region of radius $a$ near a corner of such a system as 
shown below.

\vspace{0.1in}
\centerline{\includegraphics[width=4in]{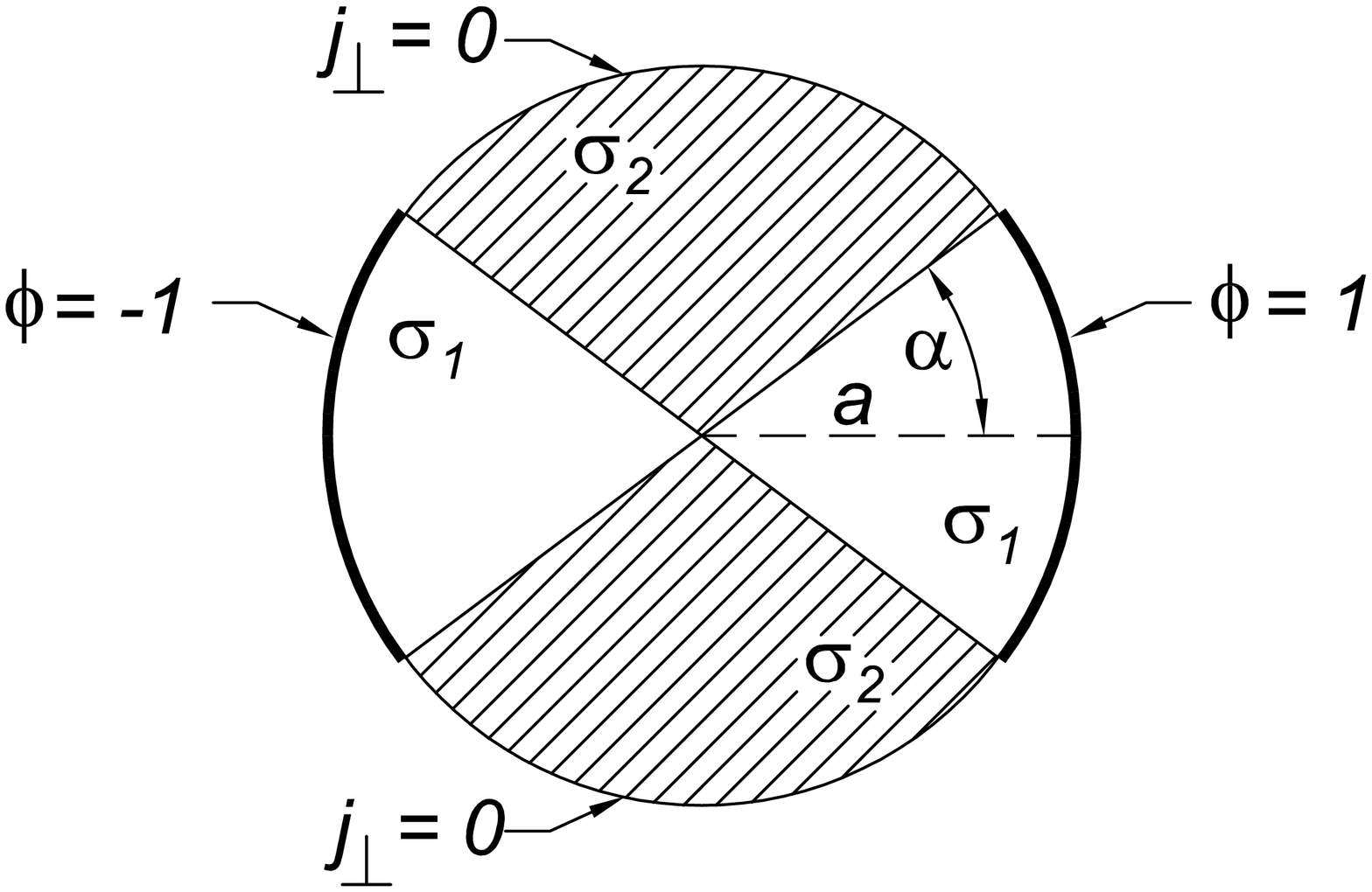}}

Phase 1, with electrical conductivity $\sigma_1$, occupies the ``bowtie'' region
of angle $\pm \alpha$, while phase 2, with conductivity $\sigma_2 \ll \sigma_1$,
occupies the remaining region.

Deduce the approximate form of lines of current density {\bf j} when a
background electric field is applied along the symmetry axis of phase 1.
What is the effective conductivity $\sigma$ of the system, defined by the
relation $I = \sigma \Delta\phi$ between the total current $I$ and the potential
difference $\Delta\phi$ across the system?

It suffices to consider the case that the boundary arc $(r = a,\abs{\theta} <
\alpha)$ is held at electric potential $\phi = 1$, while the arc 
$(r = a,\pi - \alpha < \abs{\theta} < \pi)$ is held at electric potential 
$\phi = -1$, and no current flows across the remainder of the boundary.

\medskip

Hint: When $\sigma_2 \ll \sigma_1$, the electric potential is
well described by the leading term of a series expansion.

\section{Solution}

The series expansion approach is unsuccessful in treating the full problem of a ``checkerboard" array of two phases if those phases meet in
sharp corners as shown above.  However, an analytic form for the electric potential of a two-phase (and also a four-phase) checkerboard can be obtained
using conformal mapping of certain elliptic functions \cite{Obnosov}. 
If the regions of
one phase are completely surrounded by the other phase, rather lengthy
series expansions for the potential can be given \cite{Keda}.
The present problem is based on work by Grimvall \cite{Grimvall} and
Keller \cite{Keller}.

In the steady state, the electric field obeys $\nabla \times {\bf E} = 0$,
so that {\bf E} can be deduced from a scalar potential $\phi$ via ${\bf E} 
= - \nabla \phi$.  The steady current density obeys $\nabla \cdot {\bf j} = 0$,
and is related to the electric field by Ohm's law, ${\bf j} = \sigma {\bf E}$.
Hence, within regions of uniform conductivity, $\nabla \cdot {\bf E} = 0$ and
$\nabla^2 \phi = 0$.  Thus, we seek solutions to Laplace's equations in the
four regions of uniform conductivity, subject to the stated boundary conditions
at the outer radius, as well as the matching conditions that $\phi$, 
$E_\parallel$, and $j_\perp$ are continuous at the boundaries
between the regions.

We analyze this two-dimensional problem in a cylindrical coordinate system
$(r,\theta)$ with origin at the corner between the phases and $\theta = 0$
along the radius vector that bisects the region whose potential is unity at
$r = a$.  The four regions of uniform conductivity are labeled $I$, $II$,
$III$ and $IV$ as shown below.

\vspace{0.1in}
\centerline{\includegraphics[width=4in]{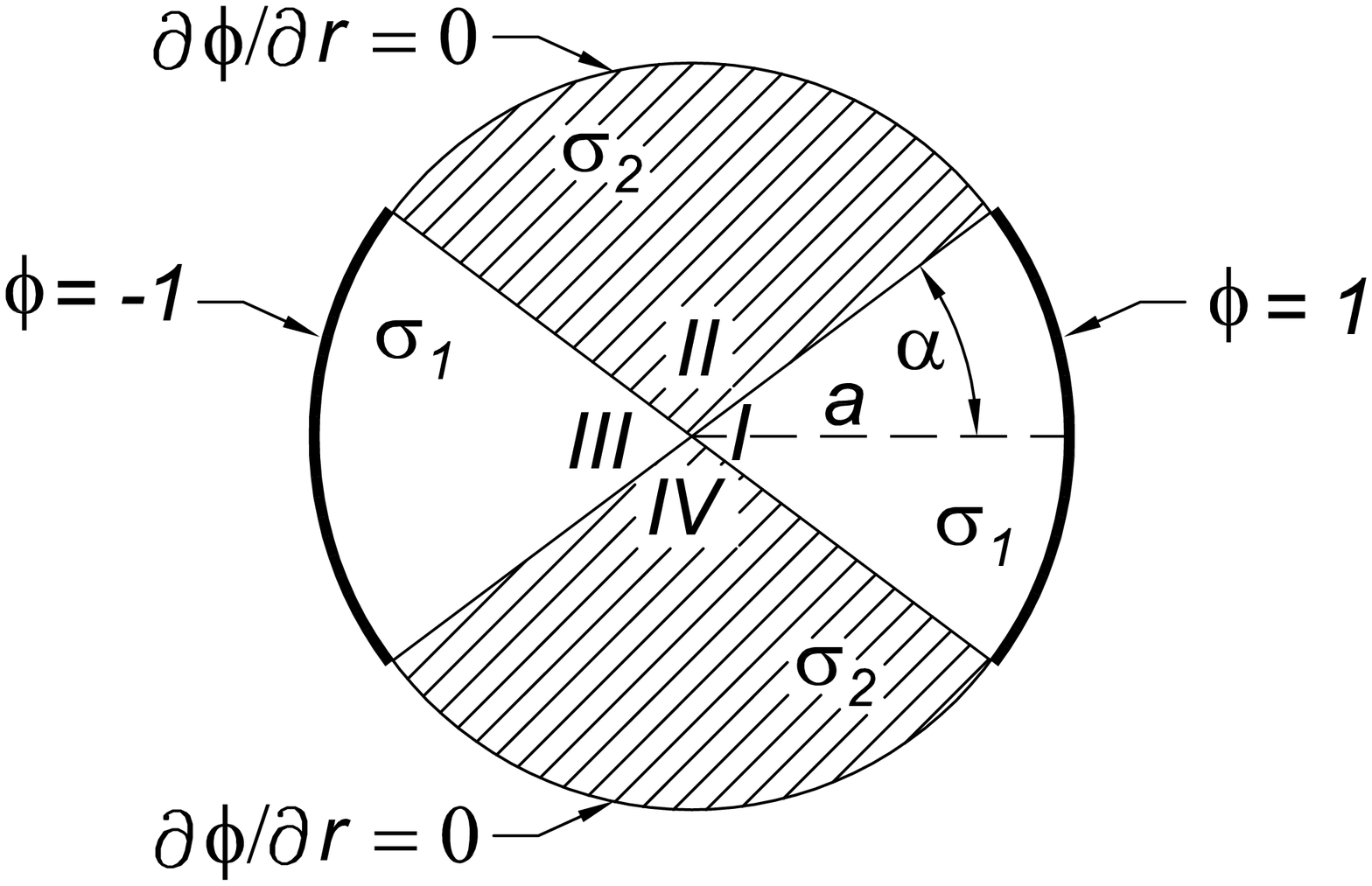}}

Since ${\bf j}_\perp = j_r = \sigma E_r = - \sigma {\partial \phi / \partial r}$
at the outer boundary, the boundary conditions at $r = a$ can be written
\begin{eqnarray}
\phi_I(r = a) & = & 1,
\label{s1} \\
{\partial \phi_{II}(r = a) \over \partial r} & = & 
{\partial \phi_{IV}(r = a) \over \partial r} = 0,
\label{s2} \\
\phi_{III}(r = a) & = & -1.
\label{s3}
\end{eqnarray}
Likewise, the condition that $j_\perp = j_\theta = \sigma E_\theta
= - (\sigma / r) \partial \phi / \partial \theta$ is continuous at the 
boundaries between the regions can be written
\begin{eqnarray}
\sigma_1 {\partial \phi_{I}(\theta = \alpha) \over \partial \theta} & = & 
\sigma_2 {\partial \phi_{II}(\theta = \alpha) \over \partial \theta}\, ,
\label{s4} \\
\sigma_1 {\partial \phi_{III}(\theta = \pi - \alpha) \over \partial \theta} & = & 
\sigma_2 {\partial \phi_{II}(\theta = \pi - \alpha) \over \partial \theta}\, ,
\label{s5} \\
&\etc& \nonumber
\end{eqnarray}

From the symmetry of the problem we see that
\begin{equation}
\phi(-\theta) = \phi(\theta),
\label{s6}
\end{equation}
\begin{equation}
\phi(\pi -\theta) = - \phi(\theta),
\label{s7}
\end{equation}
and in particular $\phi(r = 0) = 0 = \phi(\theta = \pm \pi / 2$).

We recall that two-dimensional solutions to Laplace's equations in cylindrical
coordinates involve sums of products of $r^{\pm k}$ and $e^{\pm ik \theta}$,
where $k$ is the separation constant that in general can take on a sequence
of values.  Since the potential is zero at the origin, the radial function is
only $r^k$.  The symmetry condition (\ref{s6}) suggests that the angular
functions for region $I$ be written as $\cos k\theta$, while the
symmetry condition (\ref{s7}) suggests that we use $\sin k(\pi/2 - \abs{\theta})$
in regions $II$ and $IV$ and $\cos k (\pi - \theta)$ in region $III$.  
That is, we consider the series expansions
\begin{eqnarray}
\phi_I & = & \sum A_k r^k \cos k \theta,
\label{s8} \\
\phi_{II} = \phi_{IV} & = & \sum B_k r^k \sin k \left( {\pi \over 2} -
\abs{\theta} \right)\, ,
\label{s9} \\
\phi_{III} & = & -\sum A_k r^k \cos k (\pi -\theta).
\label{s9a}
\end{eqnarray}

The potential must be continuous at the boundaries between the regions, which
requires
\begin{equation}
A_k \cos k \alpha = B_k \sin k \left( {\pi \over 2} -\alpha \right)\, .
\label{s10}
\end{equation} 
The normal component of the current density is also continuous across these
boundaries, so eq.~(\ref{s4}) tells us that
\begin{equation}
\sigma_1 A_k \sin k \alpha = 
\sigma_2 B_k \cos k \left( {\pi \over 2} -\alpha \right)\, .
\label{s11}
\end{equation} 
On dividing eq.~(\ref{s11}) by eq.~(\ref{s10}) we find that
\begin{equation}
\tan k \alpha = 
{\sigma_2 \over \sigma_1} \cot k \left( {\pi \over 2} -\alpha \right)\, .
\label{s12}
\end{equation} 
There is an infinite set of solutions to this transcendental equation.
When $\sigma_2 / \sigma_1 \ll 1$ we expect that only the first term in
the expansions (\ref{s8})-(\ref{s9}) will be important, and in this case
we expect that both $k \alpha$ and $k (\pi / 2 - \alpha)$ are small.  Then 
eq.~(\ref{s12}) can be approximated as
\begin{equation}
k \alpha \approx 
{\sigma_2 / \sigma_1 \over k ( {\pi \over 2} -\alpha )}\, ,
\label{s13}
\end{equation} 
and hence
\begin{equation}
k^2 \approx 
{\sigma_2 / \sigma_1 \over \alpha ( {\pi \over 2} -\alpha )} \ll 1.
\label{s14}
\end{equation}
Equation~(\ref{s10}) also tells us that for small $k \alpha$,
\begin{equation}
A_k \approx B_k k \left( {\pi \over 2} -\alpha \right)\, .
\label{s15}
\end{equation} 

Since we now approximate $\phi_I$ by the single term $A_k r^k \cos k\theta
\approx A_k r^k$, the boundary condition (\ref{s1}) at $r = a$ implies that
\begin{equation}
A_k \approx {1 \over a^k}\, ,
\label{s16}
\end{equation} 
and eq.~(\ref{s15}) then gives
\begin{equation}
B_k \approx {1 \over k a^k ( {\pi \over 2} -\alpha )} \gg A_k.
\label{s17}
\end{equation} 
The boundary condition (\ref{s2}) now becomes
\begin{equation}
0 =  kB_k a^{k - 1} \sin k \left( {\pi \over 2} -\theta \right)
\approx {k ({\pi \over 2} -\theta) \over  a ( {\pi \over 2} -\alpha )}\, ,
\label{s18}
\end{equation}
which is approximately satisfied for small $k$.  

So we accept the first terms of eqs.~(\ref{s8})-(\ref{s9a}) as our solution,
with $k$, $A_k$ and $B_k$ given by eqs.~(\ref{s14}), (\ref{s16}) and (
\ref{s17}).

In region $I$ the electric field is given by
\begin{eqnarray}
E_r & = & - {\partial \phi_I \over \partial r} 
\approx - k {r^{k - 1} \over a^k} \cos k\theta
 \approx - k {r^{k - 1} \over a^k}\, ,
\label{s19} \\
E_\theta & = & - {1 \over r} {\partial \phi_I \over \partial \theta} 
\approx  k {r^{k - 1} \over a^k} \sin k\theta
\approx  k^2 \theta {r^{k - 1} \over a^k}\, .
\label{s20}
\end{eqnarray}
Thus, in region $I$, $E_\theta / E_r \approx k \theta \ll 1$, so the electric
field, and the current density, is nearly radial.
In region $II$ the electric field is given by
\begin{eqnarray}
E_r & = & - {\partial \phi_{II} \over \partial r} 
\approx - k {r^{k - 1} \over k a^k ( {\pi \over 2} -\alpha )} 
\sin k \left( {\pi \over 2} -\theta \right)
 \approx - k {r^{k - 1} \over a^k} {{\pi \over 2} -\theta \over 
{\pi \over 2} -\alpha}\, ,
\label{s21} \\
E_\theta & = & - {1 \over r} {\partial \phi_{II} \over \partial \theta} 
\approx  k {r^{k - 1} \over k a^k ( {\pi \over 2} -\alpha )}
\cos k \left( {\pi \over 2} -\theta \right)
\approx  {r^{k - 1} \over a^k ( {\pi \over 2} -\alpha )}\, .
\label{s22}
\end{eqnarray}
Thus, in region $II$, $E_r / E_\theta \approx k (\pi / 2 - \theta) \ll 1$, 
so the electric field, and the current density, is almost purely azimuthal.

The current density {\bf j} follows the lines of the electric field {\bf E},
and therefore behaves as sketched below:

\vspace{0.1in}
\centerline{\includegraphics[width=4in]{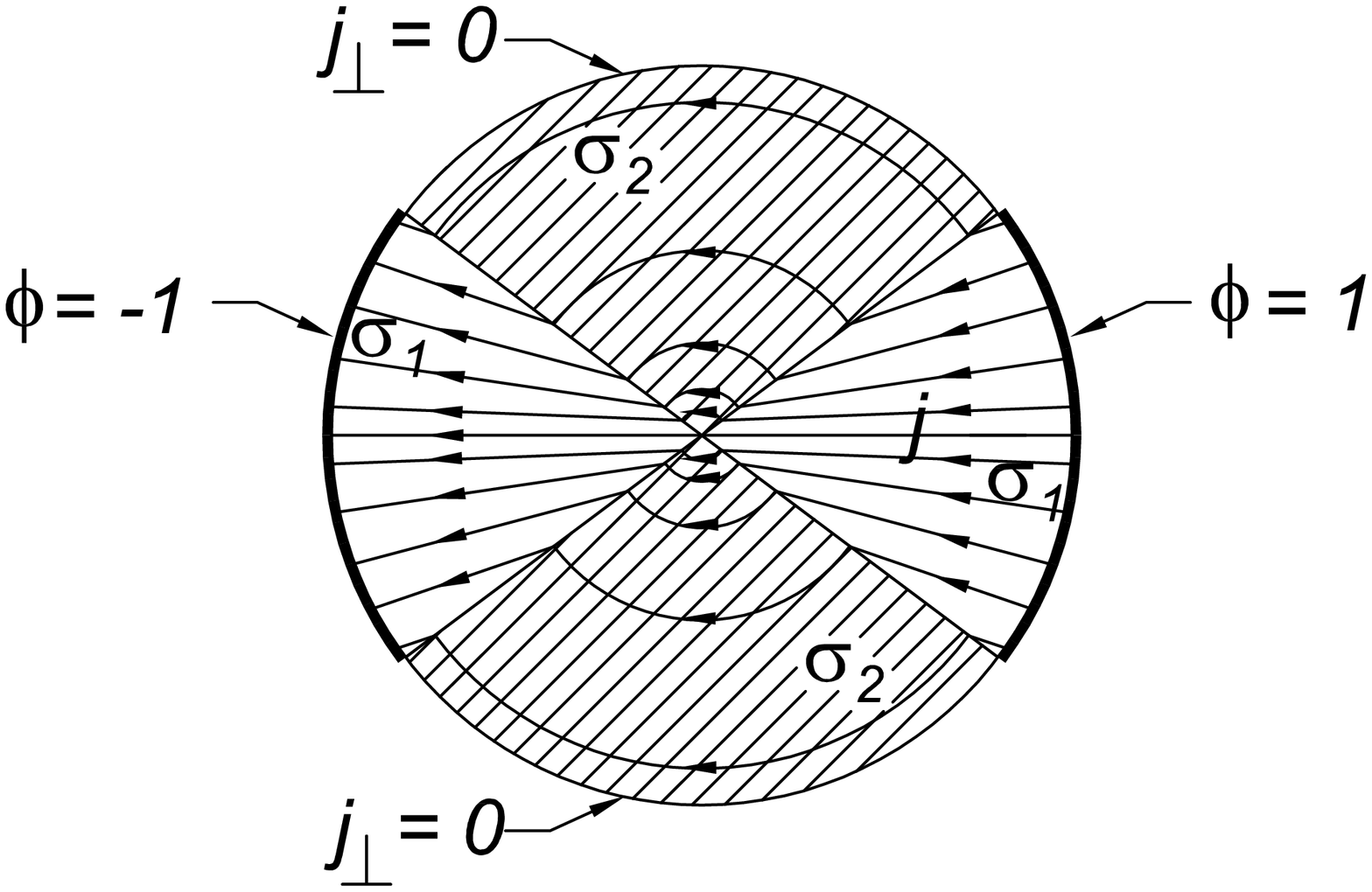}}

The total current can be evaluated by integrating the current density at
$r = a$ in region $I$:
\begin{equation}
I = 2 a \int_0^\alpha j_r d\theta
= 2 a \sigma_1 \int_0^\alpha E_r(r=a) d\theta
\approx - 2 k \sigma_1 \int_0^\alpha  d\theta
= - 2 k \sigma_1 \alpha
= - 2 \sqrt{\sigma_1 \sigma_2 \alpha \over {\pi \over 2} - \alpha}\, .
\label{s23}
\end{equation}
In the present problem the total potential difference $\Delta\phi$ is -2, so
the effective conductivity is
\begin{equation}
\sigma = {I \over \Delta \phi} 
= \sqrt{\sigma_1 \sigma_2 \alpha \over {\pi \over 2} - \alpha}\, .
\label{s24}
\end{equation}

For a square checkerboard, $\alpha = \pi / 4$, and the effective conductivity
is $\sigma = \sqrt{\sigma_1 \sigma_2}$.   It turns out that this result is
independent of the ratio $\sigma_2 / \sigma_1$, and holds not only for the corner region studied here but for the entire checkerboard array \cite{Keller64}.

\end{document}